\documentclass[11pt]{revtex4-2}
\usepackage[utf8]{inputenc}
\usepackage{graphicx}
\usepackage[colorlinks=true,citecolor=blue]{hyperref}

\begin{document}
\title{Hadron spectroscopy at STCF}
\author{Feng-Kun Guo$^{1,4}$, Haiping Peng$^{2}$, Ju-Jun Xie$^{3,4}$, Xiaorong Zhou$^{2}$ \\
{\it\footnotesize
$^{1}$ Institute of Theoretical Physics, Chinese Academy of Science, Beijing 100190, China\\
$^{2}$ University of Science and Technology of China, Hefei 230026, China\\
$^{3}$ Institute of Modern Physics, Chinese Academy of Sciences, Lanzhou 730000, China\\
$^{4}$ University of Chinese Academy of Science, Beijing 100049, China\\
}
\footnotesize{(on behalf of the STCF working group)}
}
\date{March 2022}

\maketitle

\section{Introduction}

Despite that quantum chromodynamics (QCD), the theory of strong interaction, has colorful quarks and gluons as its basic degrees of freedom, all fundamental particles participating the strong interaction that can be directly detected in experiments are colorless or color-singlet hadrons. This phenomenon is called color confinement. Because of that, the study of hadron spectroscopy is essential in improving our understanding of the nonperturbative regime of QCD and to reveal the mechanism of color confinement. The high-luminosity electron-positron collision machine under discussion, Super $\tau$-Charm Facility~(STCF), can play an essential role in hadron spectroscopy by discovering new hadron resonances and measuring properties of known hadrons with an unprecedented precision. In the following, we will discuss briefly the topics on hadron spectroscopy that will be investigated at STCF.

\section{Physics landscape}

\subsection{Highly excited charmonia and charmonium-like states}

\begin{figure*}[t]
	\centering
	\includegraphics[width=\textwidth]{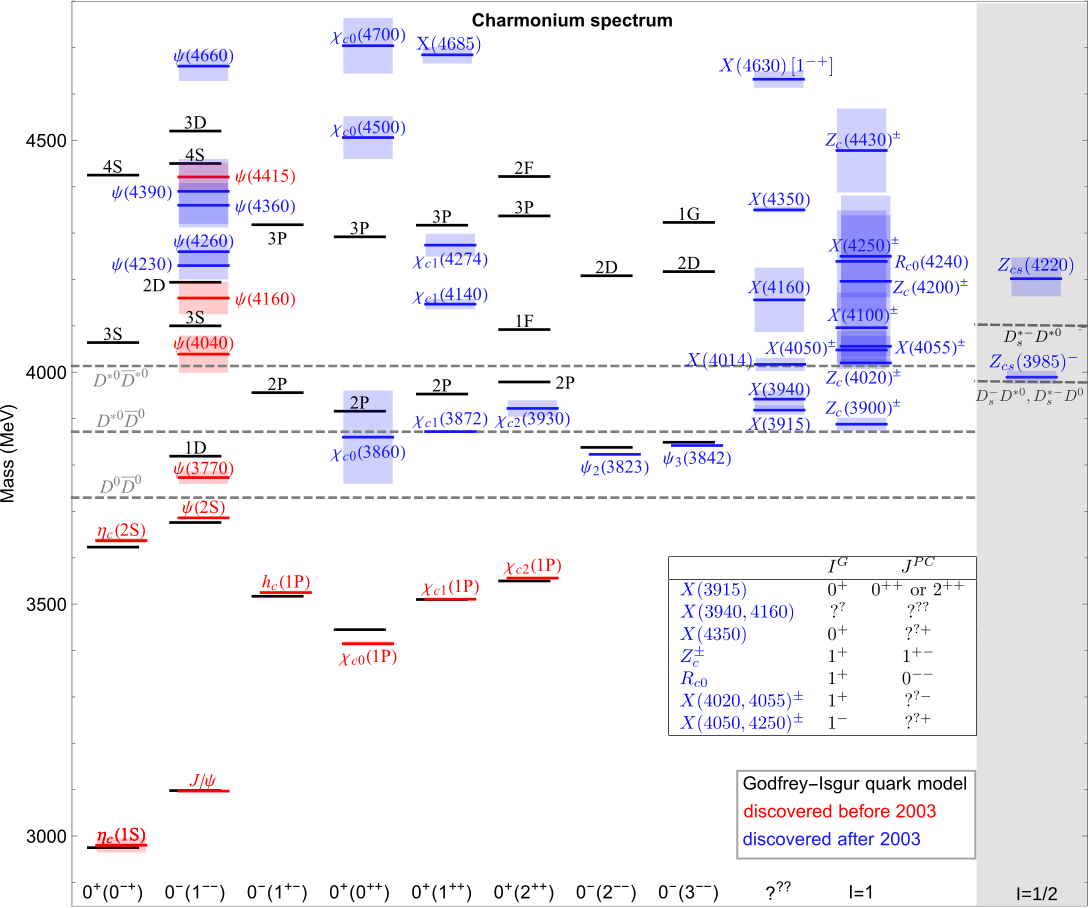}
	\vspace{0cm}
	\caption{The mass spectrum of charmonia and charmonium-like states in comparison with the predictions from the Godfrey-Isgur quark model~\cite{Godfrey:1985xj}.
		\label{fig:ccbarspec}}
\end{figure*}

Charmonium states being bound states of a charm and an anticharm quark were supposed to be well described by nonrelativistic potential quark models~\cite{Eichten:1979ms,Godfrey:1985xj}. This was the case before  2003.
Since the discovery of the $X(3872)$, also known as $\chi_{c1}(3872)$, by Belle in 2003~\cite{Belle:2003nnu}, there have been a large number of new resonance(-like) structures observed in the charmonium mass region by various  experiments, including BESIII, BaBar, Belle, CDF, D0, ATLAS, CMS and LHCb (see {\it e.g.} Refs.~\cite{Chen:2016qju,Hosaka:2016pey,Lebed:2016hpi,Esposito:2016noz,Guo:2017jvc,Olsen:2017bmm,Karliner:2017qhf,Yuan:2018inv,Liu:2019zoy,Brambilla:2019esw,Guo:2019twa,Yuan:2021wpg} for recent reviews). They are shown in Fig.~\ref{fig:ccbarspec} in comparison with the predictions of the Godfrey--Isgur quark model~\cite{Godfrey:1985xj}. Most of them have peculiar features that deviate from quark model expectations:
\begin{itemize}
    \item Masses are a few tens of MeV away from the quark model predictions for charmonia with the same quark numbers, and cannot be easily accommodated in the charmonium spectrum predicted in quark model. Examples include the $X(3872)$, $Y(4260)$, $Y(4360)$, see Fig.~\ref{fig:ccbarspec}.
    \item All of the $XYZ$ states are above or at least in the vicinity of open-charm thresholds. For those above thresholds, one would expect them to dominantly decay into open-charm channels because of the Okubo-Zweig-Iizuka rule. However, many of them have only been seen as peaks in final states of a charmonium and light mesons/photon. For instance, several positive-$C$-parity resonant structures were observed in the $J/\psi\phi$ final states~\cite{CDF:2009jgo,Belle:2009rkh,D0:2013jvp,CMS:2013jru,BaBar:2014wwp,LHCb:2021uow} and no signal of them was reported in open charm channels.
    \item Charged structures were observed, such as the  $Z_c(3900)$~\cite{BESIII:2013ris,Belle:2013yex}, $Z_c(4020)$~\cite{BESIII:2013ouc}, $Z_c(4200)$~\cite{Belle:2014nuw} and $Z_c(4430)$~\cite{Belle:2007hrb}. More recently, charged $Z_{cs}$ structures with explicit strangeness were also reported~\cite{Ablikim:2020hsk,Aaij:2021ivw}. If hidden charm hadronic resonances are the main origin for producing these structures, they must contain at least four quarks, making explicitly exotic multiquark states beyond the conventional quark model.
    % \item The narrow ones, such as the $X(3872)$, $Z_c(3900)$ and $Z_{cs}(3985)$, are close to the thresholds of a pair of charm hadrons.
\end{itemize}
Because of these features, $XYZ$ particles are thus excellent candidates of exotic hadrons. Models that are being discussed include hadronic molecules, tetraquarks, hadro-charmonia and hybrids in this context and have been searched for decades (see {\it e.g.} Refs.~\cite{Chen:2016qju,Hosaka:2016pey,Lebed:2016hpi,Esposito:2016noz,Guo:2017jvc,Olsen:2017bmm,Karliner:2017qhf,Yuan:2018inv,Liu:2019zoy,Brambilla:2019esw,Guo:2019twa} for recent reviews). However, so far it is not clear yet how the whole spectrum of charmonium-like structures can be understood. Classifying the charmonium-like structures may lead to insights into the confinement mechanism. 
To achieve such a goal, the role of kinematical effects such as triangle singularities~\cite{Guo:2019twa} needs to be further clarified in producing some of the structures.
This can be done in light of data with higher statistics at future high luminosity machines. 
%%%%%%%%%%%%%%
%%%%%%%%%%%%%%

\begin{figure*}[tbhp]
	\centering
	\includegraphics[width=\textwidth]{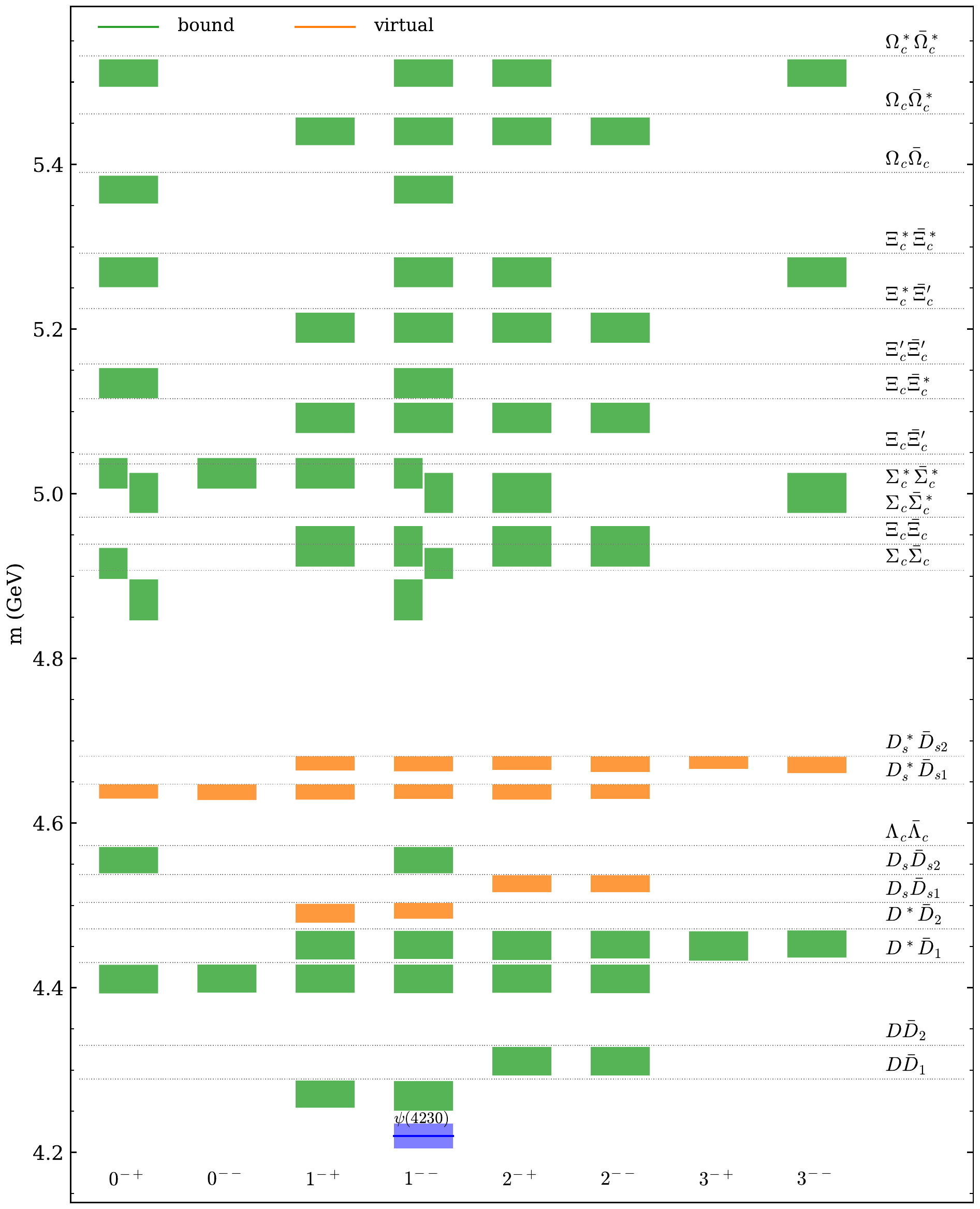}
	\vspace{0cm}
	\caption{The mass spectrum of hidden-charm hadronic molecules with negative parity predicted in a light-vector-meson-exchange model~\cite{Dong:2021juy}.
		\label{fig:hmprediction}}
\end{figure*}

Most of the narrow $XYZ$ structures are close to the thresholds of a pair of open-flavor hadrons. This salient feature may be understood as a general consequence of the $S$-wave attraction between the corresponding hadron pair~\cite{Dong:2020hxe}. Consequently, they are good candidates of hadronic molecules. 
A general pattern of the mass spectrum of negative-parity hadronic molecular from a survey considering only the exchange of light vector mesons for the hadron-hadron interaction is presented in Fig.~\ref{fig:hmprediction}~\cite{Dong:2021juy}.
One sees that a large number of the predicted states are above 4.8~GeV. The STCF running in the region from 5 to 7 GeV can play a unique role in searching for hidden-charm resonances in this energy region, in particular for these with $J^{PC}=1^{--}$ quantum numbers.
%%%%%%%%%%%%%%
%%%%%%%%%%%%%%

% \subsection{Singly-charmed mesons and baryons}

\subsection{Fully-charm tetraquarks}

The spectrum of fully-heavy tetraquarks can probe the interplay between perturbative and nonperturbtive QCD and thus are interesting objects to be studied. 
Theoretically the existence of fully-charm tetraquarks have been predicted in the above-6-GeV region, covered by STCF, since long~\cite{Iwasaki:1975pv,Chao:1980dv}. 
While whether the ground state $c c \bar{c} \bar{c}$ is below the double- $J / \psi$ or double- $\eta_{c}$ threshold is uncertain, the low-lying $c c \bar{c} \bar{c}$ states are expected to decay dominantly into final states containing a pair of charm and anti-charm hadrons via annihilating a $c \bar{c}$ pair into a gluon, and the widths are of the order of 100~MeV~\cite{Chao:1980dv,Anwar:2017toa}. Excited states with a mass well above 6.2~GeV threshold can also easily decay into $J / \psi J / \psi$, if allowed by the quantum numbers. The LHCb measurement of the double- $J / \psi$ invariant mass spectrum in semi-inclusive processes of $p p$ collisions shows clear evidence for the existence of such states~\cite{LHCb:2020bwg}. Searching for fully-charm tetraquarks in final states other than charged leptons is difficult at hadron colliders due to the huge background, and the STCF can play a rather unique role.

\subsection{Light hadrons}

The light quark sector is more complicated and challenging than that of heavy quarks, but an understanding of light quark systems is an absolute necessity to claim a full understanding hadrons. For intermediate and long-distance phenomena such as hadron properties, the full complexity of QCD emerges, which makes theoretical predictions limited and quite model dependent. In recent years, significant progresses have been made in lattice QCD~(LQCD). LQCD calculations predict the main features of the meson spectrum and provide indications for multiplets with exotic quantum numbers~\cite{Dudek:2010wm}. LQCD calculations
consistently show that the $J^{PC}=1^{-+}$ nonet is the lightest hybrid. Currently, there are three experimental candidates for a light $1^{-+}$ hybrid, which are all isovectors. There are experimental and interpretational issues surrounding them~\cite{Meyer:2010ku, Meyer:2015eta}. Recently, a isoscalar  $1^{-+}$ state, $\eta_1(1855)$ is observed in $J/\psi\to\gamma\eta\eta^{\prime}$ at BESIII~\cite{BESIII:2022riz, BESIII:2022qzu}, which may provide critical information about the $1^{-+}$ hybrid nonet. 

The self-interacting nature of gluons remains one of the most fascinating characters of QCD. A direct observation of glueball states will be the ultimate validation of low energy QCD. While quenched lattice simulations can make clear predictions for the glueball spectrum~\cite{Bali:1993fb,Morningstar:1999rf,Chen:2005mg}, the mixing of schemes in the presence of light quarks is a challenge~\cite{Crede:2008vw}. The production property
suggests the $f_0(1710)$ is largely overlapped with a scalar glueball~\cite{Sarantsev:2021ein, Rodas:2021tyb, Jin:2021vct}. However, the scalar meson sector is the most complex one and the interpretation of the state’s nature and
nonet assignments are still very controversial.

Baryons are the basic building blocks of our world. Since baryons represent the simplest
system in which all the three colors of QCD neutralize into colorless objects. Many fundamental issues in baryon spectroscopy
are still not well understood~\cite{Klempt:2009pi, Crede:2013kia}. Most important among them is the problem of missing resonances: in quark models based on approximate 
avor SU(3) symmetry it is expected that resonances form multiplets; many excited states are predicted which have
not been observed. The possibility of new, as yet unappreciated, effective symmetries could be addressed with the accumulation of more data. The new symmetries may not have obvious relation with QCD, just like nuclear shell model and collective motion model. In addition to baryons made of up and down quarks, the search for hyperon resonances remains an important challenge. Some of the lowest excitation resonances have not yet been experimentally established, which are necessary to establish the spectral pattern of hyperon resonances.

\section{Experimental landscape}
Data with unprecedented high statistics and various reaction mechanism obtained from existing (Belle/Belle II, BESIII, COMPASS, GlueX, LHCb) and future facilities (PANDA) will extend our knowledge of the complexity of QCD. BESIII/BEPCII is
undergoing upgrade for the next decade.
In addition to their main objective of studying CP-violation, Belle II and LHCb will continue to yield important hadron
spectroscopy results. Furthermore, experiments whose primary
goal is not the study of hadron spectroscopy, like ATLAS and CMS at LHC, have been able to make
significant contribution to this field. PANDA is designed for high-precision studies of the hadron spectrum at center-of-mass energies ($\sqrt{s}$) between
2 and 5.5 GeV. It is scheduled to start data taking with full setup in the next few years.

\section{STCF project}
The proposed STCF in China is a symmetric electron-positron collider designed to provide $e^{+}e^{-}$ interactions at $\sqrt{s}=2.0\sim7.0$ GeV. The peaking luminosity is expected to be of $0.5\times10^{35}$~cm$^{-2}$s$^{-1}$ or higher at $\sqrt{s}=4.0$~GeV. 
The STCF would leave space for higher luminosity upgrades and for the implementation of a longituinal polarized $e^{-}$ beam in a phase-II project. The STCF, operated at the transition interval between non-perturbative QCD and perturbative QCD,  will be one of the crucial
precision frontier for exploring  the nature of non-perturbative strong interactions. The experimental data will provide essential information to study QCD dynamics of confinement through the study of hadron spectroscopy.
The energy region covers the pair production thresholds for the recently discovered doubly charmed baryon, the exotic charmonium-like $XYZ$ states, charmed baryons, charm mesons, $\tau$-leptons, and all of the  hyperons.

The STCF is expected to deliver more than 1~ab$^{-1}$ data per year.
Year-long runs will produce data samples containing $\sim3\times10^{12}$ $J/\psi$ and $\sim5\times10^{11}$ $\psi(3686)$ events for in depth explorations of light hadron physics. At $\sqrt{s}=4.23$~GeV, the STCF will function as an ``$XYZ$-meson factory", producing $\sim 1\times10^{9}$ $Y(4260)$, $\sim1\times10^{8}$ each of $Z_{c}(3900)$ and $Z_{c}(4020)$, and
$\sim 5\times10^{6}$ $X(3872)$ events per year, enabling precision Argand plots measurement, studies of rare (including, non-hidden charm)
decays, precise mass and width measurements, {\it etc.}. 
Comparing with Belle II and LHCb experiments, the STCF has following unique
features 1) the threshold kinematics, 2) low multiplicity of final-state particles, 3) well-identied initial state, 4) full event reconstruction technique providing the best possible signal-to-background ratio for the processes with finnal-state neutrals.

The STCF detector, a state-of-the-art $4\pi$-solid-angle particle detector
operating at a high luminosity collider, is a general-purpose  detector.
It incorporates a tracking system composed of an inner tracker and main drift chamber, a particle identification system, an electromagnetic calorimeter, a super-conducting solenoid, and a muon detector at the outmost.
To fully exploit the physics opportunities and cope with the high luminosity, the STCF is designed with following requirements: (nearly) $4\pi$ solid angle coverage for both charged and neutral particles, and uniform response for these final states; excellent momentum and angular resolution for charged particles, with $\sigma_{p}/p=0.5\%$ at $p=1$~GeV/c; high resolution of energy and position reconstruction for photons, with $\sigma_{E}/E\approx2.5\%$ and $\sigma_{\rm pos}\approx 5$~mm at $E=1$~GeV;
superior particle identification ability ($e/\mu/\pi/K/p/\gamma$ and other neutral particles) and high detection efficiency for low momentum/energy particles; precision luminosity measurement; tolerance to high rate/background environment.

\section{Highlights}

\subsection{Highly excited charmonia and charmonium-like states}
STCF will be an ideal factory producing charmonium and charmonium-like states. Within the designed luminosity ($0.5\times 10^{35}$~cm$^{-2}$s$^{-1}$) and energy region ($2\sim 7$ GeV), an enormous amount of charmonium(-like) vector states ($J^{PC}=1^{--}$) can be produced with very low background environments. The states with the other quantum numbers will be reached via hadronic or radiative transitions. A systematic study, based on the unprecedented statistics, will be performed to search for undiscovered states and rare decays of the known ones, as well as to explore the properties of these states with high precision. It will extend the current studies on this filed significantly.

Current experiments, such as BEPCII/BESIII, B factories and LHCb have contributed a lot and will provide more exciting results of the charmonium-(like) states. However, these experiments take obvious limitations by themselves. At BESIII, studies of the vector states are mainly via scan strategy and the states of other quantum numbers are via radiative/hadronic transitions. Its peak luminosity is less than $1\times 10^{33}$~cm$^{-2}$s$^{-1}$ above $4$~GeV, that conflicts with the desire of a finer scan to study the $Y$ states, even unprecedented precise measurements of $Y(4230)$ have been achieved at BESIII. Its maximum c.m.~energy is about $5$ GeV, that results in that the observed non-vector states at BESIII are only $X(3872)$, $Z_c(3900)$, $Z_c(4020)$, and $Z_{cs}(3985)$. It is also very difficult to search for hidden-charm penta-quark states and impossible for full-charm tetra-quark states at BESIII. At B factories, the ISR processes and B decays are two main production modes that are utilized to study the $XYZ$ states. However, the relatively low selection efficiency of the ISR method somewhat counteracts the advantage of high luminosity of the B factories. And the B decays also suffer from the maximum energy problem due to the mass of B meson, and the corresponding analysis methods are usually very complex since amplitudes analysis is required to extract the physics information from multiple final states. At hadron collision experiments, B decay is a major production method of $XYZ$ states too. It is accompanied with direct production mode that suffers from high environment backgrounds problem. Compared with the current experiments, the large energy region, high luminosity, and clean environment of STCF make it a unique experiment in the studies of charmonium(-like) states .

To maximize the output for the charmonium(-like) states at STCF, there are some principles should be kept in mind. 1) Do not separate the studies of charmonium and $XYZ$ states. Due to their simpler internal structure and precise measurements below the open charm threshold, the charmonium states have been understood much better than the $XYZ$ states. It provides a foundation to the discoveries and observations of the $XYZ$ states, {\it i.e.} the $XYZ$ states are exotic because they have deviated from the expectation of conventional charmonia. Furthermore, some charmonium sates have same quantum numbers and similar masses to the observed $XYZ$ states, such as the 2P charmonium states and the $X$ states, and there are possible mixing between them. So the studies of the two kinds of states, conventional and exotic, are complementary to each other, and must be performed simultaneously and systematically. 2) Search for the multiplets of the observed states by considering the symmetries, such as angular momentum, isospin, flavor SU(3), and so on. 3) Disentangle the kinematic effects from the dynamic effects by precise measurements on the energy-dependence of the line-shapes, that will be promised by the high luminosity at STCF. It will be a direct way to clarify that some observed structures are really resonant or just kinematic effects such as cusp or singularity. 4) Pay special attentions to the energy regions that are close to the thresholds of some S-wave open-charm channels, since they are where most of the known exotic $XYZ$ states have been observed. Based on these four principles, a practical data taking strategy at STCF, on purpose of the charmonium(-like) studies, would be, at the beginning, a fine scan from $3700$~MeV to $7000$~MeV with a step of $10$~MeV and integrated luminosity of $1~\mathrm{fb}^{-1}$. Latter, it would be followed by larger data samples with several ab$^{-1}$ at each specific energy points, depending on the scan results.

With previously mentioned data samples, STCF will have, but not limited to, the following opportunities on the studies of the charmonium(-like) states.

\begin{itemize}
\item With the large statistical samples above $\sqrt{s} = 4$~GeV, a much better studies of the vector charmonium(-like) states can be performed at STCF. For example, the mass and width of $Y(4230)$ will be measured with significant improvement, as well as more decays modes including hidden-charm and open-charm final states. Its connection to the other $X$ and $Z$ states~\cite{Zhu:2021vtd}, such as $X(3872)$ and $Z_c(3900)$, will be studied systematically, and its nature will be revealed. Accompanying with these vector states, some other $Z_c$ states, such as $Z_c(4020)$ and $Z_{cs}(3985)$, will be searched and measured with the productions via emitting a pion or kaon meson by the vector states. The line-shapes of the $Z$ states will be measured precisely to unveil the effects of singularities~\cite{Wang:2013cya,Pilloni:2016obd,Yang:2020nrt}.
    
\item At present, $X(3872)$ is the only $PC=++$ states that is observed via transitions from $Y$ states, the others are all from $B$ decays. Limited by the maximum energy, studies of $X$ states at BESIII are mainly via radiative transitions. With $\sqrt{s}>4.7$ GeV, hadronic transitions from $Y$ states are possible and these processes should take larger production rates. For example, processes of $e^+ e^- \to \omega X$ and $e^+ e^- \to \phi X$ can be utilized to study $X(3915)$, $\chi_{c0}(3860)$, $\chi_{c2}(3930)$, and other $J^{++}$ states. The processes of $e^+ e^- \to \rho X$ can be utilized to study the spin partner of $Z_c$ states (similar to that of $Z_b$ proposed in Ref.~\cite{Bondar:2011ev}), {\it i.e.} the charged neutral isospin vector states with quantum numbers $(0,1,2)^{++}$ that can decay into $\pi^+ \pi^- J/\psi$ final states. At STCF, these $J^{++}$ states will be studied thoroughly.
    
\item The $J^{+-}$ states, such as the spin singlet P-wave states, can be studied via $e^+ e^- \to \eta X$, $\eta' X$, and $\pi^0 X$ at STCF systematically.
    
\item The upper limit energy of BEPCII is just beyond the threshold of $\Lambda_c^+ \bar{\Lambda}_c^-$. With $\sqrt{s}>5$~GeV, at STCF one can study the molecular states composed by charmed baryon pair or charmed meson pair~\cite{Cao:2019wwt,Dong:2021juy}.
    
\item With the data samples taken at $\sqrt{s}$ of $5 \sim 7$~GeV, at STCF the hidden charm penta-quark states, $P_c$ resonances, can be searched for via processes of $e^+ e^- \to p \bar{p} J/\psi$, $\Lambda_c \bar{D}^{(*)} \bar{p}$, $\Sigma^{(*)}_c \bar{D}^{(*)} \bar{p}$ {\it etc.}~\cite{Ma:2008gq,Shen:2016tzq}. Similarly, the penta-quark states containing both hidden charm and (hidden or open) strange quark components, $P_{cs}$ resonances, can be searched for via processes of $e^+ e^- \to \Lambda \bar{\Lambda} J/\psi$, $\Sigma \bar{\Sigma} J/\psi $, and $ \Xi \bar{\Xi} J/\psi$. Of course, channel $e^+ e^- \to \Omega \bar{\Omega} J/\psi$ is interesting for penta-quark states that only contain charm and strange quarks.
    
\item STCF will be unique on the production of $e^+ e^-$ annihilation to double charmonia with data at $\sqrt{s}$ of $6 \sim 7$~GeV~\cite{Iwasaki:1975pv,Chao:1980dv,Chao:1980dv,Anwar:2017toa}. LHCb has observed the evidence of full charm tetra-quark state via the invariant mass distribution of $J/\psi J/\psi$~\cite{LHCb:2020bwg}, but the performance of its detectors make any measurements on non-leptonic final states very difficult. At STCF, processes of $e^+ e^- \to \eta_c J/\psi$, $J/\psi \chi_{cJ}$ will be utilized to search for the full charm tetra-quark states and compared with theoretical predictions.
    
\item Searching for charmonium-like hybrid states will be another interesting topic at STCF. The most promising one would be the lowest energy state with the exotic quantum number $J^{PC} = 1^{-+}$, that cannot be a conventional $q\bar{q}$ state and its mass is predicted to be $4.1 \sim 4.3$~GeV by  LQCD~\cite{HadronSpectrum:2012gic}. Other excited hybrid states with quantum numbers $(0,1,2)^{-+}$ and $1^{--}$ are targets too. At $\sqrt{s}>4.5$ GeV, these hybrid states can be searched for via radiative transitions from highly excited charmonium states such $\psi(4S)$ and $\psi(5S)$.
    
\item There are still many excited charmonium states, that are predicted by potential model~\cite{Godfrey:1985xj} or  LQCD~\cite{HadronSpectrum:2012gic}, have not been observed till now. At STCF, with large statistics, a systematic search for these missing states can be performed. For example, the 1D multiplet states include spin triplet $1^3D_{1,2,3}$ and spin singlet $1^1D_2$ with quantum numbers $(1,2,3)^{--}$ and $2^{-+}$, respectively. At present, the $1^{--}$ state is assigned to the well known $\psi(3770)$, $2^{--}$ is assigned to $\psi_2(3823)$, $3^{--}$ is reported by LHCb as $\psi_3$ recently. However, the $2^{-+}$ state, called $\eta_{c2}$, has not been observed yet, even it is expected to take narrower width since it cannot decay into open charm final states. The resonance $\eta_{c2}$ may be produced via a radiative transition from $\psi(4040)$ that can be produced at BESIII too, while the luminosity of BESIII seems too limited to produce enough $\psi(4040)$ for an observation of $\eta_{c2}$~\cite{Asner:2008nq}. At STCF, the state $\eta_{c2}$ may be searched for via its decays into $\gamma h_c$, $\pi \pi \chi_{c1}$, or $\pi^+ \pi^- \pi^0 J/\psi$. With the extended energy region compared with BESIII, STCF can search for other highly excited charmonium states such as 5S/6S states, and even F and G states. For these searches, precise measurements on the cross sections of open charm final states, such as $D^{(*)}\bar{D}^{(*)}$, $D^{(*)}_s\bar{D}_s^{(*)}$, and $D^{(*)}\bar{D}^{(*)}\pi \pi$, are important. Furthermore, same analysis technology of these measurements can be applied to search for highly excited D meson states, that would be candidates of single charm tetra-quark states.

\item In addition to the spectrum, at STCF some M1 radiative transition processes, such as $\psi(2S) \to \gamma \eta_c(2S)$, $\eta_c(2S) \to \gamma J/\psi$, and $h_c \to \gamma \chi_{c0}$, as well as other rare E1 transitions can be measured. Precise measurements on the total width of the charmonium states and their partial widths of radiative, leptonic, and two photons decays are important, since these widths can be calculated by quark potential model~\cite{Barnes:2005pb} and  LQCD within clearly defined and simple frames. Comparison between experimental and theoretical results will be inspiring and then is highly desired.

\end{itemize}

\subsection{Gluonic excitations}

At STCF, a year of operation will provide $\sim3\times10^{12}$ $J/\psi$ and $\sim5\times10^{11}$ $\psi(3686)$ events at their peaking cross section for exploring light hadron physics. 

Glueballs and hybrid states may show their traces in
some more confirmed ways. The production rates of the $f_0(1710)$ in the gluon-rich processes $J/\psi\to\gamma KK/\pi\pi/\eta\eta$ are one order of
magnitude larger than those of the $f_0(1500)$~\cite{Sarantsev:2021ein, Rodas:2021tyb, Jin:2021vct}. In addition, The suppressed
decay rate of the $f_0(1710)$ into $\eta\eta^{\prime}$ ~\cite{BESIII:2022qzu} lends further support to
the hypothesis that the $f_0(1710)$ has a large overlap with the
ground state scalar glueball~\cite{Brunner:2015oga}. Measurements of electromagnetic couplings to glueball candidates would
be extremely useful for the clarification of the nature of these states. The radiative transition rates of a
relatively pure glueball would be anomalous relative to the expectations for a conventional $q\bar{q}$ state. The
dilepton decay modes of the light unflavored mesons give a deeper insight into meson structure, allowing
to measure transition form factors at the time-like region. A glueball should have suppressed couplings
to $\gamma\gamma$, which can be measured at STCF.

The $\eta_1(1855)$ is the first observed candidate of isoscalar $1^{-+}$ hybrid~\cite{BESIII:2022riz,BESIII:2022qzu,Qiu:2022ktc,Chen:2022qpd}. Further studies with more production mechanisms and decay modes will help clarify the nature of the $\eta_1(1855)$. Production property in flavor filter reactions ( {\it e.g.} $J/\psi\to V \eta_1$, where V stands for $\omega$ or $\phi$) can play an important role in unraveling the quark content of the $\eta_1(1855)$. Searching for the decay mode of $K_1 K$ and three body decays will be crucial to discriminate molecule hypothesis for the $\eta_1(1855)$~\cite{Dong:2022cuw,Zhang:2019ykd}. In addition, more precise study on the isoscalar $1^{-+}$ $\eta_1(1855)$, combined with previous and also future measurements of the isovector $\pi_1$ states, will provide critical clue of searching for other partners of hybrid supermulitplets. 

\subsection{Nucleon and hyperon resonances}

The lowest lying excited baryon in the strangeness $S=-1$ and isospin $I = 0$ sector is $\Lambda^*(1405)$. Its spin-parity quantum numbers are: $J^P = 1/2^-$. Before the quark model, $\Lambda^*(1405)$ was predicted as meson-baryon molecule composed of one kaon and one nucleon in 1959 and it was discovered later in 1961. With the development of quark model in the 1960s, the $\Lambda^*(1405)$ can be also described as an excited state of a $uds$ system. Until now, the nature of $\Lambda^*(1405)$ resonance is still unclear~\cite{ParticleDataGroup:2020ssz}. With the high luminosity, the proposed STCF is expected to produce more data samples of charmonium and charmed states, their decay processes can be used to explore the internal structure of $\Lambda^*(1405)$ and its isospin one partner $\Sigma^*(1400)$ state, such as $\chi_{c0}\to \bar{\Lambda}\Lambda(1405)\to \bar{\Lambda}\Sigma \pi$~\cite{Liu:2017hdx}, $\Lambda_c\rightarrow\pi^+\pi^0\pi^0\Sigma^0$~\cite{Dai:2018hqb}, $\chi_{c0}\to \bar\Sigma\Sigma\pi$~\cite{Wang:2015qta}, $\Lambda^+_c \to \eta \pi^+ \Lambda$~\cite{Xie:2017xwx}, and $\Lambda^+_c \to \pi^+ \pi^0 \pi^-\Sigma^+$~\cite{Xie:2018gbi}. 

In the classical quark models, the excited resonances are described as excitation of individual constituent quarks. For nucleon resonances, the lowest spatial excited state is expected to be $N^*(1535)$ with $J^P = 1/2^-$. But, it is heavier than $N^*(1440)$ with $J^P =1/2^+$ and also $\Lambda^*(1405)$, which has a strange quark. This problem can be easily solved qualitatively within the five-quark picture that one should consider five-quark components in these excited baryon states~\cite{Helminen:2000jb,Liu:2005pm,Zou:2007mk}, or taking these baryon resonances as meson-baryon dynamically generated states~\cite{Kaiser:1995eg,Oller:2000fj,Inoue:2001ip,Hyodo:2002pk}.

For these nucleon resonance with mass around 2~GeV, many of them which were predicted by the classical quark model, have not been observed so far~\cite{Capstick:2000qj}. These resonances may couple to $N\eta'$ and $N\phi$ channels which have higher mass threshold. Their couplings to $\gamma N$ and $\pi N$ channels may be too weak to be observed by the current $\gamma N$ and $\pi N$ data. However, the experimental data for the $N\eta'$ and $N\phi$ channels are very poor since their total cross sections are very small. This makes it quite challenging to obtain high-precision data for these channels. Nevertheless, it is worth investigating these channels because the high mass of the $\eta'$ and $\phi$ meson allows us to explore the high-mass region of the nucleon excitation spectrum, where a lot of missing resonances are predicted. With the platform of STCF, there will be striking opportunities in this field and a vast amount of new and diverse data can be expected in the near future. This will provide a fruitful environment for researchers and will lead to an outstanding impact in the understanding of QCD in the nonperturbative region.

Furthermore, the low-lying $\Omega$ excited states with negative parity could play a unique role for studying the five-quark picture~\cite{An:2014lga}. Because all three valence quarks inside a $\Omega^*$ state are strange ones, the five-quark configuration with a light $q\bar{q}$ pair must play a very special role in its properties. In 2018, a new $\Omega^*(2012)$ was observed by Belle~\cite{Belle:2018mqs}. It is the first $\Omega$ excited state with favored negative parity~\cite{Belle:2019zco,Liu:2019wdr,Lin:2019tex}. Yet, it is important to remark that (i) the $\Omega(2012)$ resonance was only observed by the Belle collaboration and (ii) most of its properties, such as decay fractions and spin, are not determined yet.

The $\Omega(2012)$ is naturally accommodated as a dynamically generated state from the coupled
channels interactions of the $\bar{K}\Xi(1530)$ and $\eta \Omega$ in $S$-wave~\cite{Ikeno:2020vqv,Lu:2020ste,Liu:2020yen}. In this picture, its sin-parity are $3/2^-$, and the current experimental data could be well reproduced with different sets of model parameters. Further work and more precise experimental data on its decay properties are needed~\cite{Zeng:2020och,Belle:2021gtf}.

To complete light baryon spectra~\cite{Hyodo:2020czb} and establish the low-lying hyperons with partial wave analysis, the STCF will provide an excellent place for investigating those excited baryon resonances.

\section{Prospects}
In comparison with Belle II and LHCb, STCF can contribute significantly in studying hadron spectroscopy, especially the charmonium-like states, because STCF has the advantage of clearly defined initial and final state properties and  full event reconstruction technique providing the best possible signal-to-background ratio for the processes with neutral final states. STCF is the unique place to systematically study of the vast majority of the charmonium family in ultimate high precision.
In addition, its high-statistics data sets of charmonia
provide a gluon-rich environment to further investigate light QCD exotics and new excited baryons.

\bibliography{refs}
\end{document}